\documentclass[twocolumn,showpacs,preprintnumbers,amsmath,amssymb,prb]{revtex4}

\usepackage{graphicx}
\usepackage{bm}

\begin{document}

\title{Temperature and field dependence of MgB$_2$ energy gaps from tunneling spectra}

\author{Mohamed H. Badr$^{1,2}$ and K.-W. Ng$^1$} 
\affiliation{
$^{1}$Department of Physics and Astronomy, University of Kentucky, Lexington, KY 40506-0055, U.S.A.\\
$^{2}$Department of Physics, Faculty of Science, Menoufiya University, Menoufiya, Egypt.}

\date{\today}

\begin{abstract}
We have synthesized MgB$_2$/Pb planar junctions to study the
temperature and field dependence of the superconducting energy gap of
MgB$_2$. The major peak occurs at $\Delta$ of about 2 meV, and this
corresponds to a $2\Delta/k_BT_c$value of 1.18.  While this is
significantly smaller than the BCS weak coupling value, there are
features in the tunneling spectra indicating the possibility of another
larger gap.  By fitting the dI/dV curves with a simple model, the
larger gap is estimated to be about 4.5 times the smaller gap.
The study of the effect of magnetic field on the junctions shows a stability only 
up to a field of 3.2T then junctions "collapsed"
into Josephson tunneling for higher fields. Estimation of the major energy gap from 
Josephson tunneling is consistent with that from quasiparticle tunneling.

\end{abstract}

\maketitle

\section{\label{sec:level1}Introduction\protect\\}
One interesting feature in the superconductivity of MgB$_2$
(T$_c$ = 40 K)\cite{Nagam} is the possibility of coexistence of two energy gaps.
This immediately leads to many intriguing questions, like whether
the two gaps follow $\Delta_{BCS}$(T) and share the same critical temperature.
There have been numerous tunneling spectroscopic studies to
investigate this phenomenon at low temperatures. Most of
these studies are performed with scanning tunneling microscope\cite{Rubio}
 or point contact\cite{Szabo}.  To provide the needed stability for temperature
and field dependence studies, we have synthesized planar tunnel
junctions\cite{Badr} on bulk MgB$_2$ with Pb as the counter electrode.

\section{Results and discussions}

As indicated in figure 1,\cite{Badr} the major peak ($\Delta_1$) in most of
our junctions occurs consistently around 2 meV. This is significantly
smaller than the BCS value.  The second gap ($\Delta_2$) appears as
a small feature at around 9 meV.  For conductance curves when Pb is
normal, we can fit the data with a simple model by mixing two BCS
density of states (with energy gaps $\Delta_1$ and $\Delta_2$) at a
ratio of C. A depairing term $\Gamma$ is introduced to account for
different depairing effects, and barrier strength Z is also
included to account for the zero bias offset by Andreev reflection\cite{Blond}.
Thermal broadening is included at the particular temperature of the
fitting. C, $\Gamma$, and Z are determined by the curve at temperature = 7.78 K
 above Pb $T_c\approx$ 7.2 K.  They have the best fitted values of 0.064,
0.95 meV, and 1.33 meV respectively.  These values are then fixed for all
subsequent fittings at higher temperatures. $\Delta_1$
and $\Delta_2$ are the only adjustable parameters for higher
temperature curves.

\begin{figure}
%h=here, t=top, b=bottom, p=separate figure page
\begin{center}\leavevmode
\includegraphics[width=0.9\linewidth]{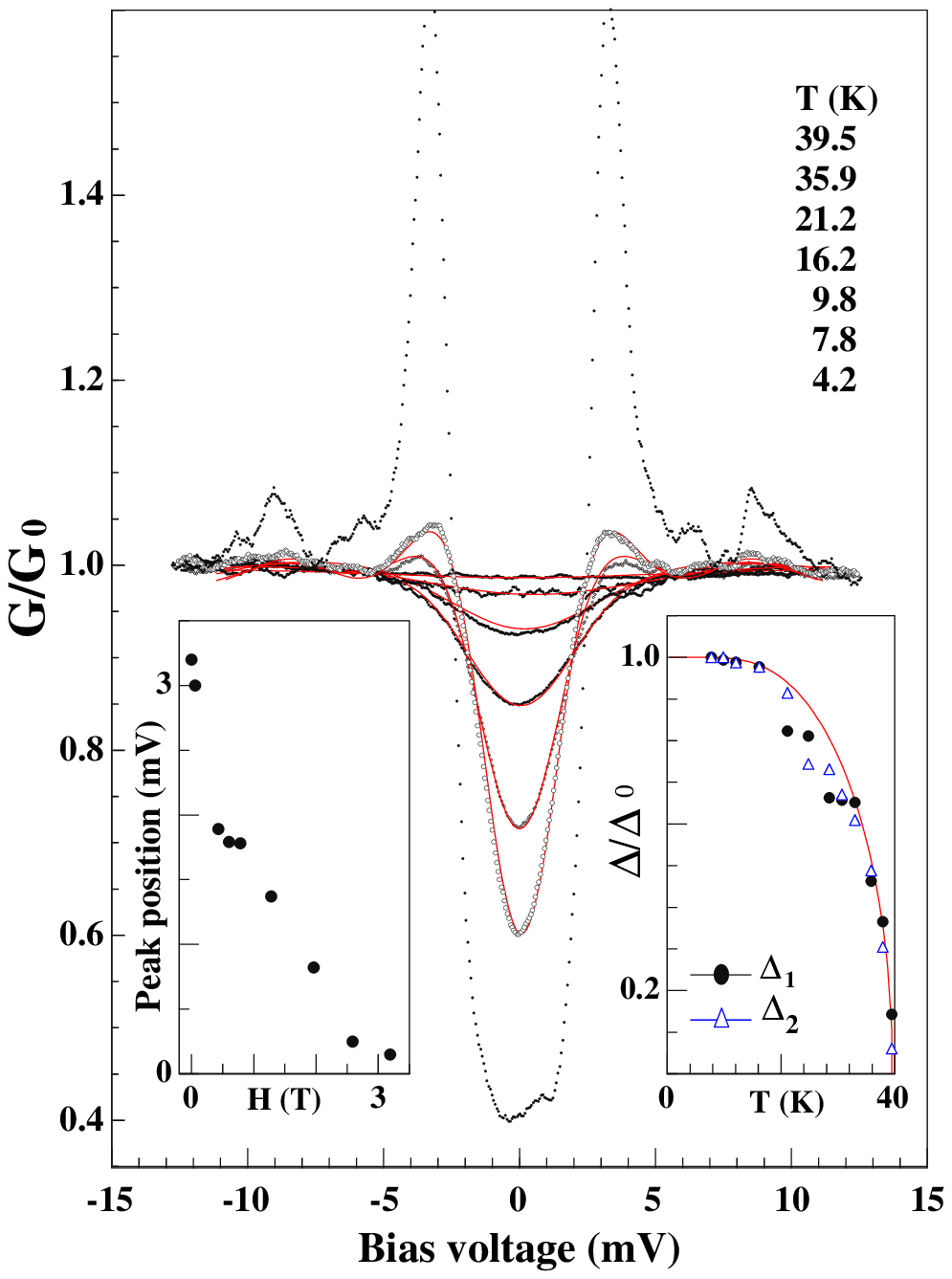}
\caption{Temperature dependence of the normalized conductance.  The lowest
curve at 4.2 K is SIS tunneling, while the others are SIN tunneling. Solid lines
are the best fitted curves from which $\Delta_1$ and $\Delta_2$ are determined.
Right insert: Temperature dependence of $\Delta_1$ and $\Delta_2$. $\Delta_0$ is
the gap value at 7.78 K.  Left insert: Field dependence of the major peak position.}
\label{Fig. 1.}\end{center}\end{figure}

The temperature dependence of the two energy gaps is shown in the
right insert of figure 1.  It is clear that both energy gaps follow a BCS-like
behavior and both gaps survive up to the bulk T$_c$ of MgB$_2$. From this
we can conclude that the commonly observed small energy gap is not
a result of surface degradation, but a true bulk property of MgB$_2$.
$\Delta_1$(0) and $\Delta_2$(0) are 1.8 meV and 8.2 meV respectively, and
the ratio $\Delta_2$/$\Delta_1$ is about 4.5 through out the temperature
range. This ratio is close to both the theoretically predicted\cite{Liu}
and experimentally suggested\cite{Bouqu} values.

The left insert in figure 1 shows the field dependence of the major
peak position at 4.2 K.  At low fields, this corresponds to the sum of
Pb gap ($\Delta_{Pb}$) and $\Delta_1$.  It can be seen that there is a
discontinuity in the peak position at the critical field of Pb ($H_c(0) \approx 0.08T)$,
when Pb ceases to be superconducting.  The peak position corresponds to $\Delta_{1}$
at all higher fields.  At a field of about 3.2 T most of our junctions "collapse"
into Josephson tunneling and the change is not
reversible.

\begin{figure}
%h=here, t=top, b=bottom, p=separate figure page
\begin{center}\leavevmode
\includegraphics[width=0.9\linewidth]{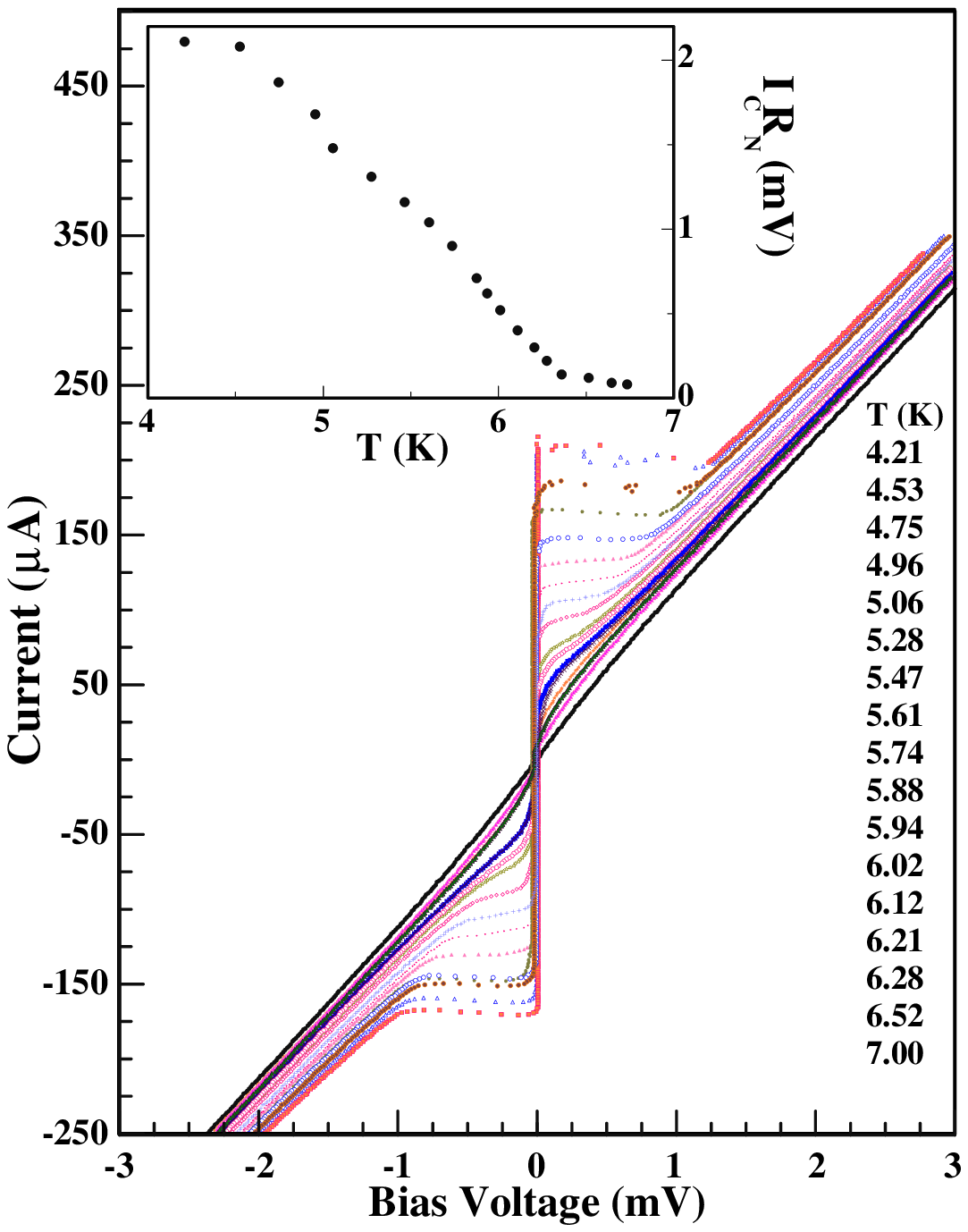}
\caption{Josephson tunneling at different temperatures.  The listed temperatures
are in the same order as the curves presented.  Insert:  $I_{c}R_{N}  vs.  T$.}
\label{Fig. 2.}\end{center}\end{figure}

The behavior of one of these Josephson junctions is shown in figure 2.
The Josephson tunneling is between MgB$_2$ and Pb.  The normal resistance
($R_{N}$) varies only very slightly with temperature. $I_{c}R_{N}$ is
estimated to be about 2.1 mV (insert, Fig. 2) for the curve at 4.2 K.  If we assume
$I_{c}R_{N} \approx \frac{\pi}{e} \frac{\Delta_{1}\Delta_{Pb}}{\Delta_{1}+\Delta_{Pb}}$
and $\Delta_{Pb} (at 4.2 K) \approx$ 1.08 meV, we can estimate $\Delta_{1}$ to be
1.75 meV.  This is consistent with our result from the quasiparticle tunneling discussed above.

This work was supported by NSF Grant No. DMR9972071. The corresponding author
would like to thank the Egyptian Ministry of Higher Education
and Scientific Research for supporting this work as well.

Corresponding author Email: mhbadr0@uky.edu

\end{document}